\documentclass[10pt,conference]{IEEEtran}
\IEEEoverridecommandlockouts
\usepackage{amsmath,stmaryrd,citesort,graphicx,mathrsfs,amssymb}

\newtheorem{theorem}{Theorem}
\newtheorem{lemma}{Lemma}
\newtheorem{definition}{Definition}
\newtheorem{remark}{Remark}

\newcommand{\set}[1]{\mathscr{#1}}
\newcommand{\XX}{\set{X}}
\newcommand{\YY}{\set{Y}}
\newcommand{\UU}{\set{U}}
\newcommand{\VV}{\set{V}}
\newcommand{\WW}{\set{W}}
\newcommand{\PP}{\set{P}}
\newcommand{\MM}{\set{M}}
\newcommand{\AAA}{\set{A}}
\newcommand{\BB}{\set{B}}
\newcommand{\CC}{\set{C}}

\newcommand{\ST}{\mathcal{A}^{*(n)}_{\epsilon}}

\newcommand{\QQ}{\mathcal{Q}}

\newcommand{\xh}{\widehat{x}}
\newcommand{\Xh}{\widehat{X}}
\newcommand{\yh}{\widehat{y}}
\newcommand{\Yh}{\widehat{Y}}

\newcommand{\wh}{\widehat{w}}

\newcommand{\RR}{\set{R}}
\newcommand{\RRo}{\set{R}_{out}}
\newcommand{\RRi}{\set{R}_{in}}

\graphicspath{{.}{./figs/}}

\begin{document}

\author{
\authorblockN{R. Timo$^{1,2}$, A. Grant$^3$, T. Chan$^3$ and G. Kramer$^4$}
\authorblockA{$^1$Department of Engineering, the Australian National University, Canberra, ACT, Australia.}
\authorblockA{$^2$Networked Systems, NICTA, Canberra Research Laboratory, ACT, Australia.}
\authorblockA{$^3$The Institute for Telecommunications Research, the University of South Australia, Adelaide, SA, Australia.}
\authorblockA{$^4$Bell Laboratories, Alcatel-Lucent, Murray Hill, NJ, USA.}
\authorblockA{roy.timo@anu.edu.au, \{alex.grant, terence.chan\}@unisa.edu.au, gkr@research.bell-labs.com.}
\thanks{NICTA is funded by the Australian Government as represented by the Department of Broadband, Communications and the Digital Economy and the Australian Research Council through the ICT Centre of Excellence program.}
\thanks{The work presented in this paper was undertaken by R. Timo while on visit at the Institute for Telecommunications Research, the University of South Australia, and Bell Laboratories, Alcatel-Lucent. R. Timo's travel was funded by student travel scholarships from NICTA, ARC Communications Research Network (ACoRN), and the University of South Australia.}}
\title{Source Coding for a Simple Network with Receiver Side Information}

\maketitle
\begin{abstract}
We consider the problem of source coding with receiver side information for the simple network proposed by R. Gray and A. Wyner in 1974. In this network, a transmitter must reliably transport the output of two correlated information sources to two receivers using three noiseless channels: a public channel which connects the transmitter to both receivers, and two private channels which connect the transmitter directly to each receiver. We extend Gray and Wyner's original problem by permitting side information to be present at each receiver. We derive inner and outer bounds for the achievable rate region and, for three special cases, we show that the outer bound is tight.
\end{abstract}

\section{Introduction}

The field of network source coding is centered on the following problem: given a noiseless communications network and a set of information sources, what is the best way to compress the output of each source for efficient and reliable transportation over the network? A solution to this type of problem needs to remove any temporal redundancy in each source, exploit any statistical correlations between different sources and optimize the use of limited channel capacities.

In network source coding, a {\em code} is a collection of rules that define how the output of each source is to be compressed, transported over the network and reconstructed. A code is said to be {\em reliable} if the output of each source can be reconstructed without error at each of its intended destinations. The {\em performance} of a reliable code is measured by the rates at which it sends data over each channel; an optimal code will send data at the smallest rates and thereby consume the least network capacity. An ordered collection of rates (one for each channel) is said to be {\em achievable} if there exists a reliable code which operates at these rates. The set of all achievable rates $\RR$ is called the {\em achievable rate region} of the network, and its lower boundary $\overline{\RR}$ provides a performance benchmark for the comparison of reliable codes.

The achievable rate region $\RR$ is known for a small ad-hoc collection of networks; for most ``real world" networks, $\RR$ is unknown~\cite{Effros-Dec-2007-A}. With the exception of~\cite{Csiszar-Mar-1980-A}, achievable rate regions have been studied on a network-by-network basis; researchers have designed and studied simple networks which isolate particular problems of interest. Two notable examples are: the separate coding of correlated sources~\cite{Slepian-Jul-1973-A}, and the sharing of a finite capacity channel between multiple users~\cite{Wyner-Jun-2002-A}. It is hoped that solutions to these simple networks will yield practical and efficient codes for larger networks.

\begin{figure}[htb]
\setlength{\unitlength}{0.001\columnwidth}
\begin{picture}(950,480)
    \put(0,0){\centerline{\includegraphics[width=1.0\columnwidth]{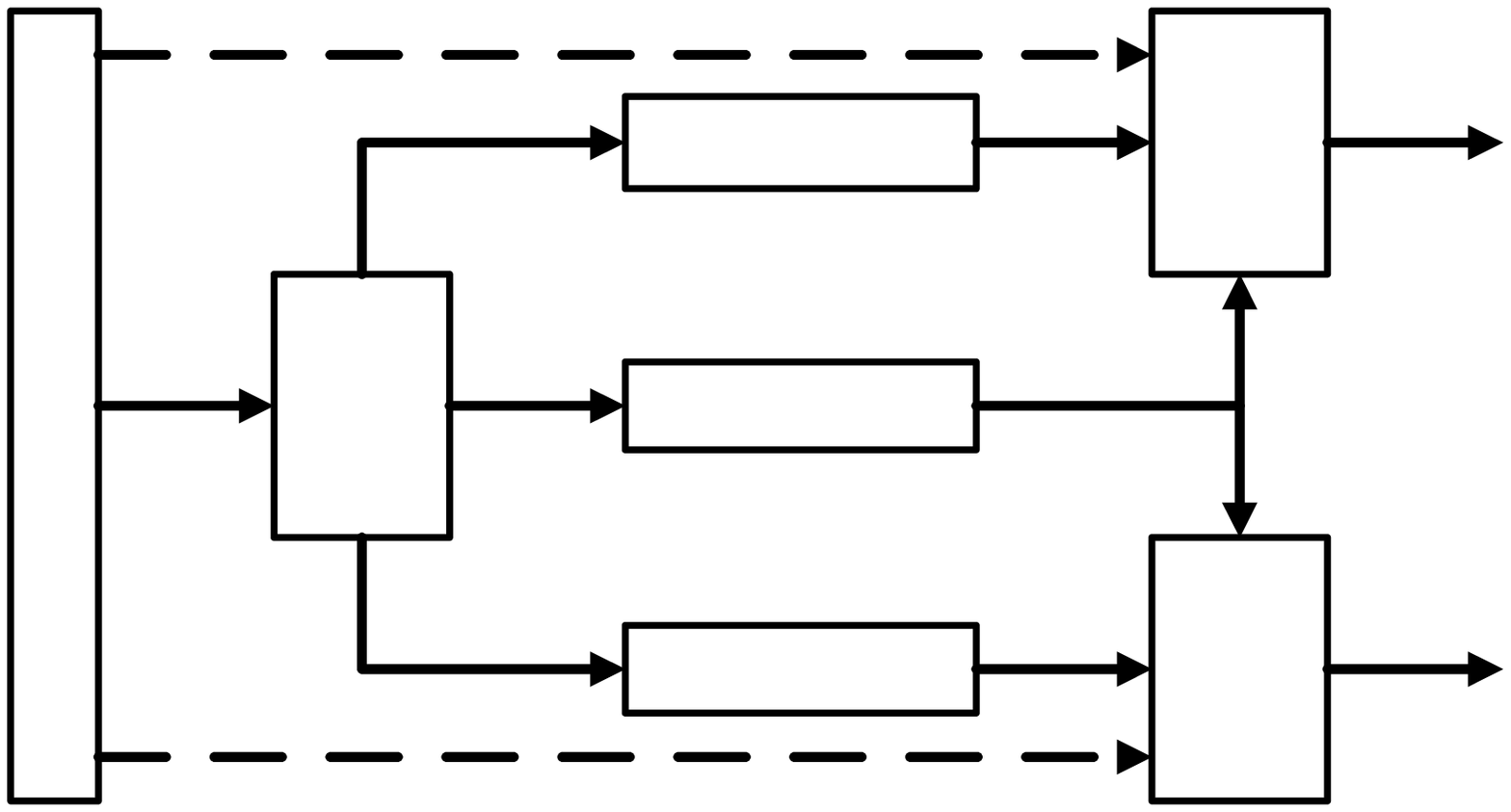}}}
    \put(60,50){\small{\rotatebox{90}{Discrete Memoryless Source}}}
    \put(130,270){\small{\rotatebox{90}{$\{(X_i,Y_i)\}$}}}
    \put(235,200){\small{\rotatebox{90}{Encoder}}}
    \put(430,400){\small{Channel 1}}
    \put(430,245){\small{Channel 0}}
    \put(430,95){\small{Channel 2}}
    \put(740,350){\small{\rotatebox{90}{Decoder}}}
    \put(740,45){\small{\rotatebox{90}{Decoder}}}
    \put(840,440){\small{$\{\Xh_i\}$}}
    \put(840,60){\small{$\{\Yh_i\}$}}
    \put(130,480){\small{$\{U_i\}$}}
    \put(130,10){\small{$\{V_i\}$}}
    \put(740,495){\small{$X$-Receiver}}
    \put(740,0){\small{$Y$-Receiver}}
\end{picture}
\caption{Figure shows the network source coding problem proposed by R. Gray and A. Wyner~\cite{Gray-Nov-1974-A}. The transmitter is connected to two receivers via three noiseless channels. The sequences $\{X_i\}$ and $\{Y_i\}$ are to be encoded at the transmitter, transported over the network and decoded at the $x$ and $y$-receivers respectively. In this paper, we study an extension of this problem where ``side information" $\{U_i\}$ and $\{V_i\}$ are present at each receiver. These additional information sources are marked with dashed lines in the figure.}\label{Sec:1:Fig:1}
\end{figure}

We study the achievable rate region $\RR$ of the network shown in Figure~\ref{Sec:1:Fig:1}. A transmitter must transport the output of two correlated sources to two receivers using three noiseless channels: a public channel which connects the transmitter to both receivers, and two private channels which connect the transmitter directly to each receiver. The achievable rate region $\set{R}$ of this network was found by R. Gray and A. Wyner~\cite{Gray-Nov-1974-A} in $1974$. They showed that an optimal code should endeavor to use the public channel to transport information common to both sources. As we will see, the intuition of this solution is lost when side information is introduced at each receiver; in particular, it is not clear how one should decompose the output of each source for transmission over the three channels.

An outline of the paper is as follows. To fix ideas, we briefly review~\cite{Gray-Nov-1974-A} in Section~\ref{Sec:2}. In Section~\ref{Sec:3}, we formally define $\RR$ for the network with side information. In Sections~\ref{Sec:4} and~\ref{Sec:5}, we derive outer and inner bounds for $\RR$ respectively. In Section~\ref{Sec:6}, we ascertain $\RR$ for one source, a degraded network and a complementary delivery network respectively. Finally, we conclude the paper in Section~\ref{Sec:7}.

\section{The Gray-Wyner Problem}\label{Sec:2}

Consider the network (without receiver side information) shown in Figure~\ref{Sec:1:Fig:1}. We denote the capacities (in bits per second) of channels $0$, $1$ and $2$ by $C_0$, $C_1$ and $C_2$ respectively. Finally, let $\XX$ and $\YY$ be finite alphabets, and let $\XX^n$ and $\YY^n$ denote their respective $n$-fold cartesian product spaces.

Suppose $\{(X_i,Y_i)\} \triangleq \{(X_i,Y_i);\ i = 1,2,\ldots\}$ is a sequence of independent and identically distributed (i.i.d.) $\XX \times \YY$ valued random variables emitted by a discrete memoryless source $\QQ_{XY}(x,y) = \text{Prob}[X = x, Y = y]$. Suppose further that the random sequence $\{(X_i,Y_i)\}$ appears at the transmitter at the rate of one per second. It is desired that the transmitter delivers a reliable reproduction $\{\Xh_i\} \triangleq \{\Xh_i;\ i = 1,2,\ldots\}$ of the sequence $\{X_i\}$ to the $x$-receiver, and a reliable reproduction $\{\Yh_i \} \triangleq \{\Yh_i;\ i = 1,2,\ldots\}$ of the sequence $\{Y_i\}$ to the $y$-receiver. Assuming no delay constraints and unlimited computational power at the transmitter and receivers, the main problem is to ascertain which channel capacity triples $(C_0,C_1,C_2)$ are both necessary and sufficient for each sequence to be reliably transported to its intended destination.

We assume the classic $n$-block source coding model where the sequence $\{(X_i,Y_i)\}$ is parsed and transported over the network in message blocks of length $n$ (for some large integer $n$). Let $(X^n,Y^n) = (X_1,Y_1),(X_2,Y_2),\ldots,(X_n,Y_n)$ denote the message at the transmitter, and let $\Xh^n = \widehat{X}_1,\widehat{X}_2,\ldots,\widehat{X}_n$ and $\Yh^n = \widehat{Y}_1,\widehat{Y}_2,\ldots,\widehat{Y}_n$ denote the reconstructed messages at the $x$ and $y$-receivers respectively.

For each $i = 0,1,2$, let $\set{M}_i = \{1,2,\ldots,|\set{M}_i|\}$ be a finite index set for use on channel $i$. A network source code is a collection of mappings $(e^{(n)},d_x^{(n)},d_y^{(n)})$, where $e^{(n)}: \set{X}^n \times \set{Y}^n \rightarrow \set{M}_0 \times \set{M}_1 \times \set{M}_2$ is the encoder at the transmitter; $d^{(n)}_x: \set{M}_0 \times \set{M}_1 \rightarrow \set{X}^n$ is the decoder at the $x$-receiver; and $d^{(n)}_y: \set{M}_0 \times \set{M}_2 \rightarrow \set{Y}^n$ is the decoder at the $y$-receiver. The transmitter encodes the pair $(X^n,Y^n)$ with three indices $(M_0,M_1,M_2) = e^{(n)}(X^n,Y^n)$ which are sent over channels $0$, $1$ and $2$ respectively. After receiving indices $M_0$ and $M_1$, the $x$-receiver reconstructs $\widehat{X}^n = d^{(n)}_x(M_0,M_1)$. Similarly, after receiving indices $M_0$ and $M_2$, the $y$-receiver reconstructs $\widehat{Y}^n = d^{(n)}_y(M_0,M_2)$. An error is said to occur if either $\Xh^n \neq X^n$ or $\Yh^n \neq Y^n$, and the code is said to operate at a rate of $(1/n) \log_2 |\set{M}_i|$ bits per source symbol on channel $i$ (for $i = 0,1,2$).

A triple of rates $(R_0,R_1,R_2)$ is said to be achievable if there exists a sequence of codes $\{(e^{(n)},d_x^{(n)},d_y^{(n)});\ n = 1,2,\ldots\}$ such that the probability of error approaches zero and $(1/n)\log|\set{M}_i|$ approaches $R_i$ (for $i=0,1,2$) as $n$ goes to infinity.

Let $\RR_{GW}$ denote the set of all achievable rate triples. It can be shown that $\RR_{GW}$ is a closed convex subset of Euclidean three space, which is completely defined by its lower boundary $\overline{\RR}_{GW}$~\cite{Gray-Nov-1974-A}:
\begin{multline*}
\overline{\RR}_{GW} \triangleq \big\{(R_0,R_1,R_2) \in \RR_{GW} : (\widehat{R}_0,\widehat{R}_1,\widehat{R}_2) \in \RR_{GW},\\ \widehat{R}_i \leq R_i\ (i = 0,1,2) \rightarrow \widehat{R}_i = R_i\ (i =0,1,2)\big\}\ .
\end{multline*}

Given $\QQ_{XY}$ and a network with capacity triple $(C_0,C_1,$ $C_2)$, the sequences $\{X_i\}$ and $\{Y_i\}$ may be reliably reconstructed at the $x$ and $y$-receivers respectively if and only if $(C_0,C_1,C_2)$ lies above $\overline{\RR}_{GW}$; thus, $\overline{\RR}_{GW}$ defines exactly those capacity triples which are both necessary and sufficient for reliable communication.

Gray and Wyner~\cite{Gray-Nov-1974-A} showed that to achieve rates $(R_0,R_1,R_2)$ which lie on the lower boundary $\overline{\RR}_{GW}$, the capacity of channel $0$ should be prioritized for use by information common to both $\{X_i\}$ and $\{Y_i\}$. Specifically, they designed a coding scheme which used an auxiliary random variable $W$ to represent the information transported over channel $0$, and they showed any $(R_0,R_1,R_2) \in \overline{\RR}_{GW}$ may be achieved by optimizing over the choice of $W$.

The formal description of $\set{R}_{GW}$ in terms of $W$ is as follow. Let $\WW$ be a finite alphabet of cardinality $|\set{W}| \leq |\set{X}||\set{Y}|+2$, and let $\PP_{GW}$ denote the family of probability functions on $\WW \times \XX \times \YY$ such that $\sum_{w} p(w,x,y) = \QQ_{XY}(x,y)$. Now, for each $p \in \PP_{GW}$, let
\begin{equation*}
\RR_{GW}^{(p)} \triangleq \left\{ (R_0,R_1,R_2)\ :
\begin{array}{lll}
                 R_0 &\geq& I_p(X,Y;W)\\
                 R_1 &\geq& H_p(X|W)\\
                 R_2 &\geq& H_p(Y|W)
\end{array}
\right\}\ ,
\end{equation*}
where $I_p(\cdot;\cdot)$ denotes mutual information and $H_p(\cdot|\cdot)$ denotes conditional entropy (with respect to $p$).

\begin{lemma}~\cite[Thm. 4]{Gray-Nov-1974-A}\label{Sec:1:Lem:1}
The achievable rate region $\set{R}_{GW}$ of the Gray-Wyner Network is given by
\begin{equation*}
\set{R}_{GW} = \left(\bigcup_{p \in \set{P}_{GW}} \set{R}_{GW}^{(p)} \right)^c\ ,
\end{equation*} 
where $(\cdot)^c$ denotes the set closure operation. 
\end{lemma}

It follows from Lemma~\ref{Sec:1:Lem:1} that $\RR_{GW}$ is completely described by a single coding scheme which makes use of an auxiliary random variable $W$. As we will see, this coding scheme extends, in a natural way, to the network with side information. Unfortunately, however, this extension does not appear to completely describe the corresponding rate region. 
\section{Extension to the Side Information Case}\label{Sec:3}
Suppose $\XX$, $\YY$, $\UU$ and $\VV$ are finite sets, and let $\XX^n$, $\YY^n$, $\UU^n$ and $\VV^n$ denote their respective $n$-fold cartesian product spaces. Suppose further that $\{(X_i,Y_i,U_i,V_i)\}$ is a sequence of i.i.d. $\set{X} \times \set{Y} \times \set{U} \times \set{V}$ valued random variables emitted by a discrete memoryless source $\QQ_{XYUV}(x,y,u,v) = \text{Prob}\big[X = x, Y = y, U = u, V = v\big]$. Finally, for each $i = 0,1,2$, let $\MM_i = \{1,2,\ldots,|\MM_i|\}$ be a finite index set for channel $i$.

As before, a source code is a collection of mappings $(e^{(n)},d_x^{(n)},d_y^{(n)})$, where $e^{(n)}: \set{X}^n \times \set{Y}^n \rightarrow \set{M}_0 \times \set{M}_1 \times \set{M}_2$ is the encoder at the transmitter; $d_x^{(n)}: \set{M}_0 \times \set{M}_1 \times \set{U}^n \rightarrow \set{X}^n$ is the decoder at the $x$-receiver; and $d_y^{(n)}: \set{M}_0 \times \set{M}_2 \times \set{V}^n \rightarrow \set{Y}^n$ is the decoder at the $y$-receiver. The transmitter encodes the pair $(X^n,Y^n)$ with indices $(M_0,M_1,M_2) = e^{(n)}(X^n,Y^n)$ which are sent over channels $0$, $1$ and $2$ respectively. After receiving indices $M_0$ and $M_1$ as-well-as side information $U^n$, the $x$-receiver reconstructs $\widehat{X}^n = d_x^{(n)}(M_0,M_1,U^n)$. Similarly, after receiving $M_0$, $M_2$ and $V^n$, the $y$-receiver reconstructs $\widehat{Y}^n = d_y^{(n)}(M_0,M_2,V^n)$.

An error occurs if either $\widehat{X}^n \neq X^n$ or $\widehat{Y}^n \neq Y^n$. Let $P_{e,x} \triangleq \text{Prob}[\Xh^n \neq X^n]$, $P_{e,y} \triangleq \text{Prob}[\Yh^n \neq Y^n]$ and $P_e \triangleq \max\{P_{e,x},\ P_{e,y}\}$.

\begin{definition}[Achievable Rate]
A rate triple $(R_0,R_1,R_2)$ is said to be achievable if, for arbitrary $\epsilon > 0$ and sufficiently large $n$, there exists a code $(e^{(n)},d_x^{(n)},d_y^{(n)})$ with parameters $(n,|\MM_0|,|\MM_1|,|\MM_2|,P_e)$ such that $P_e \leq \epsilon$ and
$(1/n) \log |\MM_i| \leq R_i + \epsilon$ for all $i = 0,1,2$. We let $\RR$ denote the set of all achievable rate triples.
\end{definition}

\section{An Outer Bound}\label{Sec:4}

Suppose $\WW$ is a finite set of cardinality $|\set{W}| \leq |\set{X}||\set{Y}| + 3$ and $\PP$ is the family of probability functions on $\WW \times \XX \times \YY \times \UU \times \VV$ such that $p(w,x,y,u,v) = p(w|x,y)p(x,y,u,v)$ and
\begin{equation*}
\QQ_{XYUV}(x,y,u,v) = \sum_{w \in \WW}p(w,x,y,u,v)
\end{equation*}
for all $p \in \PP$. Now, for each $p \in \PP$ let
\begin{multline*}
\RRo^{(p)} = \Big\{ (R_0,R_1,R_2) : \\
               \left. \begin{array}{lll}
                 R_0 &\geq& \max\big\{I_p(X,Y;W|U),I_p(X,Y;W|V)\big\} \\
                 R_0 + R_1 &\geq& \max\big\{I_p(X,Y;W|U),I_p(X,Y;W|V)\big\} \\
                        & & \quad + H_p(X|W,U), \\
                 R_0 + R_2 &\geq& \max\big\{I_p(X,Y;W|U),I_p(X,Y;W|V)\big\} \\
                        &  &\quad + H_p(Y|W,V).
               \end{array}
             \right\}
\end{multline*}

\begin{theorem}[Outer Bound]\label{Sec:4:The:1}
If $(R_0,R_1,R_2)$ is an achievable rate triple, then there exists a $p \in \PP$ such that $(R_0,R_1,R_2) \in \RRo^{(p)}$.
\end{theorem}

\subsection{Proof Outline: Theorem~\ref{Sec:4:The:1}}

We show: if $\{(e^{(n)},d_x^{(n)},d_y^{(n)})\}$ is a sequence of codes where $P_e \rightarrow 0$ as $n \rightarrow \infty$, then there exists a $p \in \PP$ such that
$((1/n)\log|\MM_0|,\ (1/n)\log|\MM_1|,\ (1/n)\log|\MM_2|\big) \in \RR^{(p)}_{out}$.

Suppose $(e^{(n)},d^{(n)}_x,d^{(n)}_y)$ is a code with $(M_0,M_1,M_2) = e^{(n)}(X^n,Y^n)$, $\widehat{X}^n = d_x^{(n)}(M_0,M_1,U^n)$ and  $\widehat{Y}^n = d_y^{(n)}(M_0,M_2,V^n)$, then
\begin{align}
\notag              &\log|\MM_0| \geq H(M_0|U^n) \geq I(X^n,Y^n;M_0|U^n) \\
\label{Sec:4:Eqn:1} & = \sum_{i=1}^n I(X_i,Y_i;M_0,X_1^{i-1},X_1^{i-1},U_1^{i-1},U_{i+1}^n|U_i) \\
\label{Sec:4:Eqn:2}              & \geq \sum_{i=1}^n I(X_i,Y_i;M_0|U_i) = \sum_{i=1}^n I(X_i,Y_i;W_i|U_i)\ ,
\end{align}
where~\eqref{Sec:4:Eqn:1} follows because $\{(X_i,Y_i,U_i,V_i)\}$ is drawn in an i.i.d. fashion and~\eqref{Sec:4:Eqn:2} follows by setting $W_i = M_0$. Similarly,
\begin{equation}\label{Sec:4:Eqn:3}
\log|\MM_0| \geq \sum_{i=1}^n I(X_i,Y_i;W_i|V_i)\ .
\end{equation}

On applying Fano's Inequality~\cite[Pg. 37]{Cover-2006-B} we get
\begin{equation}
\label{Sec:4:Eqn:4} H(X^n|M_0,M_1,U^n)  \leq H(X^n|\widehat{X}^n) \leq n \delta(P_{e},n)\ ,
\end{equation}
where $\delta(P_{e},n) \triangleq (1/n) + P_e \log |\set{X}||\set{Y}|$. Similarly, we also have that $H(Y^n|M_0,M_1,V^n) \leq n\delta(P_e,n)$.

Now consider the series of Shannon (in)equalities~\eqref{Sec:4:Eqn:5} through~\eqref{Sec:4:Eqn:12}.
\begin{figure*}[t]
\begin{align}
\label{Sec:4:Eqn:5} \log|\MM_0| + \log|\MM_1| &\geq H(M_0,M_1) =H (M_0,M_1|U^n) + I(M_0,M_1;U^n)\\
\label{Sec:4:Eqn:6}    &\geq I(X^n,Y^n;M_0,M_1|U^n) + I(M_0,M_1;U^n)\\
\label{Sec:4:Eqn:7}    &=\sum_{i=1}^n \big[ I(X_i,Y_i;M_0,M_1,X_1^{i-1},Y_1^{i-1},U_1^{i-1},U_{i+1}^n|U_i)
                            +  I(U_i;M_0,M_1,U_1^{i-1}) \big]\\
\label{Sec:4:Eqn:8}    &\geq\sum_{i=1}^n \big[ I(X_i,Y_i;M_0,M_1,U_1^{i-1},U_{i+1}^n|U_i)
                            + I(U_i;M_0) \big]\\
\label{Sec:4:Eqn:9}    &=\sum_{i=1}^n \big[ I(X_i,Y_i;M_0|U_i) + I(X_i,Y_i;M_1,U_1^{i-1},U_{i+1}^n|M_0,U_i)
                            + I(U_i;M_0) \big]\\
\label{Sec:4:Eqn:10}    &\geq\sum_{i=1}^n \big[ I(X_i,Y_i;M_0|U_i) + I(X_i;M_1,U_1^{i-1},U_{i+1}^n|M_0,U_i)
                            + I(U_i;M_0) \big]\\
\label{Sec:4:Eqn:11}    &=\sum_{i=1}^n \big[ I(X_i,Y_i;M_0|V_i) + H(X_i|M_0,U_i) - n\delta(P_{e},n) \big]\\
\label{Sec:4:Eqn:12}    &=\sum_{i=1}^n \big[ I(X_i,Y_i;W_i|V_i) + H(X_i|W_i,U_i) - n\delta(P_{e},n) \big]
\end{align}
\hrulefill
\end{figure*}
Note, \eqref{Sec:4:Eqn:7} follows because $\{(X_i,Y_i,U_i,V)\}$ is drawn in an i.i.d. fashion and~\eqref{Sec:4:Eqn:11} follows since $M_0 \minuso (X_i,Y_i) \minuso (U_i,V_i)$ forms a Markov Chain and~\eqref{Sec:4:Eqn:4}. From~\eqref{Sec:4:Eqn:10} and~\eqref{Sec:4:Eqn:12}, it respectively follows that
\begin{align*}
&\frac{1}{n} \Big( \log|\MM_0| + \log|\MM_1| \Big) \\
&\geq \frac{1}{n}\sum_{i=1}^n\Big[I(X_i,Y_i;W_i|U_i) +  H(X_i|W_i,U_i) \Big] - \delta(P_{e},n)\ ,
\end{align*}
and
\begin{align*}
&\frac{1}{n} \Big( \log|\MM_0| + \log|\MM_1| \Big) \\
&\geq \frac{1}{n}\sum_{i=1}^n\Big[I(X_i,Y_i;W_i|V_i) +  H(X_i|W_i,U_i) \Big] - \delta(P_{e},n)\ .
\end{align*}
Note, $(1/n)[\log|\set{M}_0|+\log|\set{M}_2|]$ may be bound in a similar manner. Following the time sharing principle given in~\cite[Pg. 1709]{Gray-Nov-1974-A}, we may now construct a $p \in \PP$ such that each inequality in the theorem holds as $n \rightarrow \infty$ and $P_e \rightarrow 0$. Finally, we may bound the cardinality of the auxiliary random variable $W$ using the support lemma of Ahlswede and K$\ddot{\text{o}}$rner~\cite[Lemma 3]{Ahlswede-Nov-1975-A}.

\section{An Inner Bound}\label{Sec:5}
A natural extension of the code proposed by Gray and Wyner~\cite{Gray-Nov-1974-A} yields the following inner bound for $\RR$.

Let $\WW$ and $\PP$ be defined as in Section~\ref{Sec:3}. For $p \in \PP$, let
\begin{multline*}
\RRi^{(p)} = \Big\{ (R_0,R_1,R_2) : \\
               \left. \begin{array}{lll}
                 R_0 &\geq& \max\big\{I_p(X,Y;W|U),I_p(X,Y;W|V)\big\} \\
                 R_1 &\geq& H_p(X|W,U), \\
                 R_2 &\geq& H_p(Y|W,V).
               \end{array}
             \right\}\ ,
\end{multline*}
and $\RRi = \big(\cup_{p \in \PP} \RRi^{(p)}\big)^c$.

\begin{theorem}\label{Sec:5:The:1}
$\RR \supseteq \RRi$.
\end{theorem}

\begin{remark}
If $U = V$, then $\RR = \RRi$.
\end{remark}

\begin{remark}
Suppose $(X,Y) \minuso U \minuso V$ forms a Markov chain. It can be shown that a sum rate $R_0+R_1+R_2$ is achievable if and only if $R_0+R_1+R_2 \geq H(Y|V) + H(X|Y,U)$. (See~\cite{Timo-Feb-2007-C} for the special case where $V = \text{ constant}$.) We may set $W = Y$ in Theorem~\ref{Sec:5:The:1} to achieve this sum rate.
\end{remark}

\begin{remark}
Suppose $X = Y$. Sgarro~\cite{Sgarro-Mar-1977-A} showed that the sum rate $R_0 + R_1 + R_2$ is achievable if and only if $R_0 + R_1 + R_2 \geq \max\{H(X|U),\ H(X|V)\}$. We may set $W = X = Y$ in Theorem~\ref{Sec:5:The:1} to achieve this sum rate.
\end{remark}

\begin{remark}
Suppose $U = Y$ and $V = X$. Wyner {\em et. al.}~\cite{Wyner-Jun-2002-A} showed that the sum rate $R_0 + R_1 + R_2$ is achievable if and only if $R_0 + R_1 + R_2 \geq \max\{H(X|Y),\ H(Y|X)\}$. We may set $W = (X,Y)$ in Theorem~\ref{Sec:5:The:1} to achieve this sum rate.
\end{remark}

\begin{remark}
The code, which yields the achievability of $\RRi$, is essentially a version of Heegard and Berger's ``triple rate split code" given in~\cite[Thm. 2]{Heegard-Nov-1985-A}. Indeed, we note that the problem of minimizing the sum rate $R_0 + R_1 + R_2$ is a special case of the two receiver generalized Kaspi-Heegard-Berger problem~\cite[Sec. VII]{Heegard-Nov-1985-A}.
\end{remark}

\subsection{Proof Outline: Theorem~\ref{Sec:5:The:1}}
\subsubsection{Code Construction}
Suppose $p \in \PP$. Let $R_0'$, $R_1'$ and $R_2'$ be non-negative integers whose values will be chosen later. Generate $2^{nR_0'}$ independent $w$-codewords of length $n$ by choosing symbols i.i.d. from $\WW$ according to $p_W$ (the $W$-marginal of $p$). Label the resulting code book with the index $m_0'$:
$\CC_{\WW} \triangleq \{w^n(m_0') : 1 \leq m_0' \leq 2^{nR_0'}\}$. Similarly, generate $2^{nR_1'}$ and $2^{nR_2'}$ independent $x$ and $y$-codewords using $p_X$ and $p_Y$ respectively: $\CC_{\XX} \triangleq \{x^n(m_1') : 1 \leq m_1' \leq 2^{nR_1'}\}$,  and $\CC_{\YY} \triangleq \{y^n(m_2') : 1 \leq m_2' \leq 2^{nR_2'}\}$.

Uniformly at random assign to each $w^n \in \CC_{\WW}$ a ``bin label" from the set $\MM_0 = \{1,2,\ldots,2^{\lfloor nR_0 \rfloor}\}$, and let $h_{\WW} : \CC_{\WW} \rightarrow \MM_0$ denote the induced mapping. Let $\BB_{\WW}(m_0)$ denote the set of $w$-codewords with bin label $m_0$: $\BB_{\WW}(m_0) \triangleq \{w^n \in \CC_{\WW} : h_{\WW}(w^n) = m_0 \}$, and let $\BB_{\WW}$ denote the collection of all $w$-bins. In the same way, assign one of $2^{\lfloor nR_1 \rfloor}$ and $2^{\lfloor nR_2 \rfloor}$ bin labels to each $x$ and $y$-codeword, and define $h_{\XX}$, $h_{\YY}$, $\BB_{\XX}$ and $\BB_{\YY}$.

\subsubsection{Encoding}
The encoder assumes the messages $x^n$, $y^n$, $u^n$ and $v^n$ emitted by the source are $\epsilon$-strong joint typical; that is, $(x^n,y^n,u^n,v^n) \in \ST(p_{XYUV})$. Let $E_1$ denote the event where this assumption is false. Then~\cite[Lem. 10.6.1]{Cover-2006-B}
\begin{equation}\label{Sec:5:Eqn:1}
\text{Pr}\big[E_1\big] \leq \epsilon_1(n,\XX\times\YY\times\UU\times\VV)\ ,
\end{equation}
where $\epsilon_1(n,\XX\times\YY\times\UU\times\VV) \rightarrow 0$ in $n$ for fixed $\epsilon > 0$.

The transmitter looks for a $w^n(m_0') \in \CC_{\WW}$ which is $\epsilon$-strong joint typical with $(x^n,y^n)$. If two-or-more such codewords exist, the transmitter selects the codeword with the smallest index. If no such codeword exists, an error is declared and the transmitter arbitrarily selects some $w_e^n(m_0') \in \CC_{\WW}$. Let $E_2$ denote this error event. Then~\cite[Lem. 10.6.2]{Cover-2006-B},
\begin{equation}\label{Sec:5:Eqn:2}
\text{Pr}\big[E_2\big] \leq e^{-\left(2^{nR_0'} 2^{-n(I(X,Y;W)+\epsilon_2)}\right)}\ ,
\end{equation}
where $\epsilon_2 \rightarrow 0$ as $\epsilon \rightarrow 0$ and $n\rightarrow\infty$. We assume $R_0' \geq I(X,Y;W) + \epsilon_2$, so that $\text{Pr}[E_2] \rightarrow 0$ as $\epsilon \rightarrow 0$ and $n\rightarrow\infty$. After the transmitter selects $w^n(m_0') \in \CC_{\WW}$ it sends the index $m_0 = h_{\WW}(w^n(m_0'))$ on channel $0$.

The transmitter looks for a $x^n(m_1') \in \CC_{\XX}$ such that $x^n(m_1') = x^n$. If two-or-more such codewords exist, the transmitter selects the codeword with the smallest index. If no such codeword exists, an error is declared and the transmitter arbitrarily selects some $x_e^n(m_1') \in \CC_{\XX}$. Let $E_{3,x}$ denote this error event. Then,
\begin{equation}\label{Sec:5:Eqn:3}
\text{Pr}\big[E_{3,x}\big] \leq e^{-\left(2^{nR_1'} 2^{-n(H(X)+\epsilon_{3,x})}\right)}\ ,
\end{equation}
where $\epsilon_{3,x} \rightarrow 0$ as $\epsilon \rightarrow 0$ and $n \rightarrow \infty$. Choose $R_0' \geq H(X) + \epsilon_{3,x}$ arbitrarily, so that $\text{Pr}[E_{3,x}] \rightarrow 0$ as $\epsilon \rightarrow 0$ and $n\rightarrow\infty$. The transmitter encode $y^n$ is a similar fashion, and sends $m_1 = h_{\XX}(x^n(m_1'))$ and $m_2 = h_{\YY}(y^n(m_2'))$ on channels $1$ and $2$ respectively.

\subsubsection{Decoding}

Given $m_0$ and $u^n$, the $X$-receiver looks for a unique $\wh^n \in \BB_{\WW}(m_0)$ which is jointly typical with $u^n$. If no such codeword can be found, an error is declared and the decoder arbitrarily selects some $\wh_e^n \in \BB_{\WW}(m_0)$. Let
\begin{itemize}
\item $E_{4,x}$: the codeword $w^n(m_0')$ chosen by the transmitter is not jointly typical with $u^n$, and
\item $E_{5,x}$: there are two-or-more $w$-codewords in $\BB_{\WW}(m_0)$ which are jointly typical with $u^n$.
\end{itemize}

Consider $E_{4,x}$. Since $W \minuso (X,Y) \minuso U$ forms a Markov Chain under $p$, we have that~\cite[Lem. 15.8.1]{Cover-2006-B}
\begin{equation}\label{Sec:5:Eqn:4}
\text{Pr}\big[E_{4,x}\big] \leq \epsilon_{4,x}\ ,
\end{equation}
where $\epsilon_{4,x} \rightarrow 0$ as $n\rightarrow\infty$.

Now consider $E_{5,x}$. We have that $u^n \in \ST(P_U)$. As before, the probability that a randomly generated $w$-codeword is jointly typical with $u^n$ is upper bound by $2^{-n(I(W;U)+\epsilon_{5,x})}$, where $\epsilon_{5,x} \rightarrow 0$ as $n\rightarrow\infty$. Moreover, the number of codewords in each bin is at most $2^{n(R'_0-R_0)} + \epsilon_{5',x}$, where $\epsilon_{5',x}\rightarrow0$ as $n\rightarrow\infty$~\cite[Pg. 2766]{Gastpar-Nov-2004-A}. Hence,
\begin{equation*}
\text{Pr}\big[E_{5,x}\big] \leq 2^{-n(R_0-R_0' + I(W;U)-\epsilon_{5,x})} + \epsilon_{5',x}\ .
\end{equation*}

We need $R_0-R_0' + I(W;U)-\epsilon_{5,x} \geq 0$, so that $\text{Pr}\big[E_{5,x}\big] \rightarrow 0$ as $n \rightarrow \infty$. This requires
\begin{align}
\notag               R_0 &\geq R_0' - I(W;U) + \epsilon_{5,x} \\
\label{Sec:5:Eqn:5}     &\geq I(X,Y;W) -I(W;U) + \epsilon_2 + \epsilon_{5,x} \\
\label{Sec:5:Eqn:6}     &= I(X,Y,U;W) - I(W;U) + \epsilon_2 + \epsilon_{5,x} \\
\label{Sec:5:Eqn:7}     &= I(X,Y;W|U) + \epsilon_2 + \epsilon_{5,x} \ ,
\end{align}
where~\eqref{Sec:5:Eqn:5} follows because we selected $R_0' \geq I(X,Y;W)  + \epsilon_2$, \eqref{Sec:5:Eqn:6} follows because $W \minuso (X,Y) \minuso U$ forms a Markov Chain, and~\eqref{Sec:5:Eqn:7} follows from the chain rule for mutual information. Similarly, the $y$-receiver will correctly find a $w$-codeword with high probability if $R_0 \geq I(X,Y;W|V) + \epsilon_2 + \epsilon_{5,y}$, where $\epsilon_{5,y} \rightarrow 0$ as $\epsilon \rightarrow 0$ and $n \rightarrow 0$.

Given $\wh^n$, $m_1$ and $u^n$, the $X$-receiver looks for a unique $\xh^n \in \BB_{\XX}(m_1)$ which is jointly typical with $\wh^n$ and $u^n$. If there exists two-or-more such codewords, an error is declared and the decoder arbitrarily selects some $\xh_e^n \in \BB_{\XX}(m_1)$. Let $E_{6,x}$ denote this error event. It follows that
\begin{equation}\label{Sec:5:Eqn:8}
\text{Pr}\big[E_{6,x}\big] \leq 2^{-n(R_1-R_1' + I(X;W,U)-\epsilon_{6,x})} + \epsilon_{6',x}\ ,
\end{equation}
where $\epsilon_{6,x} \rightarrow 0$ and $\epsilon_{6',x} \rightarrow 0$ as $\epsilon \rightarrow 0$ and $n\rightarrow\infty$. If $R_1 \geq H(X|W,U) + \epsilon_{6,x}$ it follows from~\eqref{Sec:5:Eqn:8} that $\text{Pr}[E_{6,x}] \rightarrow 0$ as $n \rightarrow \infty$. Similarly, the $y$-receiver will correctly find $\yh^n$ with high probability if $R_2 \geq H(Y|W,V) + \epsilon_{6,y}$, where $\epsilon_{6,y} \rightarrow 0$ as $\epsilon \rightarrow 0$ and $n \rightarrow 0$.

\section{Three Simple Networks}\label{Sec:6}

\subsection{Two Descriptions of $\RR$ when $X = Y$}

Let $\PP$ and $\RRo^{(p)}$ be defined as in Section~\ref{Sec:4}.

\begin{theorem}\label{Sec:6:The:1}
If $X = Y$, then $\set{R} = \left(\cup_{p\in\set{P}} \RRo^{(p)} \right)^c$.
\end{theorem}

Now suppose $\AAA$ and $\BB$ are finite sets of cardinalities $|\set{A}|\leq |\set{X}|+1$ and $|\set{B}|\leq |\set{X}|+1$. Let $\PP^*$ denote the family of probability functions on $\AAA \times \BB \times \XX \times \UU \times \VV$ such that $p(a,b,x,u,v) = p(a,b|x)p(x,u,v)$ and
\begin{equation*}
\QQ_{XUV}(x,u,v) = \sum_{(a,b) \in \AAA \times \BB} p(a,b,x,u,v)\ .
\end{equation*}
For each $p \in \PP^*$, let
\begin{multline*}
\RR^{(p)*} = \Big\{ (R_0,R_1,R_2) : \\
               \left. \begin{array}{lll}
                 R_0 &\geq& \max\big\{H_p(X|A,U),H_p(X|B,V)\big\} \\
                 R_1 &\geq& I_p(X;A|U), \\
                 R_2 &\geq& I_p(X;B|V).
               \end{array}
             \right\}\ .
\end{multline*}

\begin{theorem}\label{Sec:6:The:2}
If $X = Y$, then $\RR = \left(\cup_{p\in\set{P}^*} \RR^{(p)*}\right)^c$.
\end{theorem}

\subsection{$\RR$ for a Type of Degraded Network}
Let $\PP$ and $\RRo^{(p)}$ be defined as in Section~\ref{Sec:4}.
\begin{theorem}\label{Sec:6:The:3}
If $Y = (X,Z)$ and $(X,Z) \minuso U \minuso V$ forms a Markov Chain, then 
$\set{R} = \left(\cup_{p\in\set{P}} \RRo^{(p)} \right)^c$.
\end{theorem}

\subsection{$\RR$ for a Complementary Delivery Network}

Let $\PP$ and $\RRo^{(p)}$ be defined as in Section~\ref{Sec:4}.
\begin{theorem}\label{Sec:6:The:4}
If $U = Y$ and $V = X$, then $\set{R} = \left(\cup_{p\in\set{P}} \RRo^{(p)} \right)^c$.
\end{theorem}

Now suppose $\AAA$ and $\BB$ are finite sets of cardinalities $|\set{A}| \leq |\set{X}||\set{Y}| + 1$ and $|\set{B}| \leq |\set{X}||\set{Y}| + 1$. Let $\PP^{**}$ denote the set of probability functions on $\AAA \times \BB \times \XX \times \YY$ such that
\begin{equation*}
\QQ_{XY}(x,y) = \sum_{(a,b) \in \AAA \times \BB} p(a,b,x,y)
\end{equation*}
is true for all $(x,y)$ and $p \in \PP^{**}$. For each $p \in \PP^{**}$, let
\begin{multline*}
\RR^{(p)**} = \Big\{ (R_0,R_1,R_2) : \\
               \left. \begin{array}{lll}
                 R_0 &\geq& \max\big\{H_p(X|A,Y),H_p(Y|B,X)\big\} \\
                 R_1 &\geq& I_p(X;A|Y), \\
                 R_2 &\geq& I_p(Y;B|X).
               \end{array}
             \right\}\ .
\end{multline*}

\begin{theorem}\label{Sec:6:The:5}
If $U = Y$ and $V = X$, then 
$\set{R} = \left(\cup_{p\in\set{P}^{**}} \RR^{(p)**} \right)^c$.
\end{theorem}

\section{Conclusion}\label{Sec:7}
We investigated the achievable rate region $\RR$ of a simple network with side information present at each receiver. Our first theorem gave an outer bound which, for three simple networks, was shown to be equal to $\RR$. Our second result gave an inner bound which was obtained via an extension of the coding theorem given by Gray and Wyner~\cite{Gray-Nov-1974-A}. 

\bibliographystyle{IEEEtran}


\begin{thebibliography}{}
\providecommand{\url}[1]{#1}
\csname url@rmstyle\endcsname
\providecommand{\newblock}{\relax}
\providecommand{\bibinfo}[2]{#2}
\providecommand\BIBentrySTDinterwordspacing{\spaceskip=0pt\relax}
\providecommand\BIBentryALTinterwordstretchfactor{4}
\providecommand\BIBentryALTinterwordspacing{\spaceskip=\fontdimen2\font plus
\BIBentryALTinterwordstretchfactor\fontdimen3\font minus
  \fontdimen4\font\relax}
\providecommand\BIBforeignlanguage[2]{{%
\expandafter\ifx\csname l@#1\endcsname\relax
\typeout{** WARNING: IEEEtran.bst: No hyphenation pattern has been}%
\typeout{** loaded for the language `#1'. Using the pattern for}%
\typeout{** the default language instead.}%
\else
\language=\csname l@#1\endcsname
\fi
#2}}

\end{thebibliography}


\begin{thebibliography}{10}

\bibitem{Effros-Dec-2007-A}
M.~Effros, ``Network {S}ource {C}oding: {A} {P}erspective,'' \emph{IEEE Inform.
  Theory Society Newsletter}, vol.~57, no.~4, pp. 15--23, December 2007.

\bibitem{Csiszar-Mar-1980-A}
I.~Csiszar and J.~K$\ddot{\text{o}}$rner, ``Towards a {G}eneral {T}heory of
  {S}ource {N}etworks,'' \emph{{IEEE} Transactions on Information Theory},
  vol.~26, no.~2, pp. 155--165, March 1980.

\bibitem{Slepian-Jul-1973-A}
D.~Slepian and J.~Wolf, ``Noiseless {C}oding of {C}orrelated {S}ources,''
  \emph{{IEEE} Transactions on Information Theory}, vol.~19, no.~4, pp.
  471--480, July 1973.

\bibitem{Wyner-Jun-2002-A}
A.~Wyner, J.~Wolf, and F.~Willems, ``Communicating {V}ia {P}rocessing
  {B}roadcast {S}atellite,'' \emph{{IEEE} Transactions on Information Theory},
  vol.~48, no.~6, pp. 1243--1249, June 2002.

\bibitem{Gray-Nov-1974-A}
R.~Gray and A.~Wyner, ``Source {C}oding for a {S}imple {N}etwork,'' \emph{Bell
  System Technical Journal}, vol.~53, no.~9, pp. 1681--1721, Nov. 1974.

\bibitem{Cover-2006-B}
T.~Cover and J.~Thomas, \emph{Elements of {I}nformation {T}heory},
  2nd~ed.\hskip 1em plus 0.5em minus 0.4em\relax Wiley, 2006.

\bibitem{Ahlswede-Nov-1975-A}
R.~F. Ahlswede and J.~K$\ddot{\text{o}}$rner, ``{S}ource {C}oding with {S}ide
  {I}nformation and a {C}onverse for {D}egraded {B}roadcast {C}hannels,''
  \emph{{IEEE} Transactions on Information Theory}, vol.~21, no.~6, pp.
  629--637, November 1975.

\bibitem{Timo-Feb-2007-C}
R.~Timo, A.~Grant, and L.~Hanlen, ``Source {C}oding for a {N}oiseless
  {B}roadcast {C}hannel with {P}artial {R}eceiver {S}ide {I}nformation,'' in
  \emph{Proceedings {IEEE} Australian Communications Theory Workshop, AusCTW},
  Adelaide, Australia, February 2007.

\bibitem{Sgarro-Mar-1977-A}
A.~Sgarro, ``Source {C}oding with {S}ide {I}nformation at {S}everal
  {D}ecoders,'' \emph{{IEEE} Transactions on Information Theory}, vol.~23,
  no.~2, pp. 179--182, March 1977.

\bibitem{Heegard-Nov-1985-A}
C.~Heegard and T.~Berger, ``Rate {D}istortion when {S}ide {I}nformation {M}ay
  {B}e {A}bsent,'' \emph{{IEEE} Transactions on Information Theory}, vol.~31,
  no.~6, pp. 727--734, November 1985.

\bibitem{Gastpar-Nov-2004-A}
M.~Gastpar, ``{T}he {W}yner {Z}iv {P}roblem {W}ith {M}ultiple {S}ources,''
  \emph{{IEEE} Transactions on Information Theory}, vol.~50, no.~11, pp.
  2762--2768, November 2004.

\end{thebibliography}
\end{document}